\documentclass[11pt]{article}
\usepackage[utf8]{inputenc}
\usepackage[left=2.54cm, right=2.54cm, top=2.54cm, bottom = 2.54cm]{geometry}
\usepackage{xcolor} 
\usepackage{indentfirst} 
\usepackage{caption}
\usepackage{hyperref}
\usepackage{mathtools}

\usepackage{mathrsfs}
\usepackage{amsmath, amssymb, amsfonts, amsthm,rotating,booktabs,array}
\usepackage{cancel}
\usepackage{bm}
\usepackage{booktabs}
\usepackage{subfigure}
\usepackage{graphicx}
\usepackage{mathabx}
\usepackage{setspace}

\usepackage{algorithm,algpseudocode}
\algnewcommand\algorithmicinput{\textbf{Input:}}
\algnewcommand\algorithmicoutput{\textbf{Output:}}
\algnewcommand\Input{\item[\algorithmicinput]}%
\algnewcommand\Output{\item[\algorithmicoutput]}%

\makeatletter
\def\algbackskip{\hskip-\ALG@thistlm}
\makeatother
\usepackage{mathtools}
\usepackage{multicol}
\usepackage{multirow}
\usepackage{verbatim}
\usepackage{enumitem}
\usepackage{relsize}
\usepackage{listings} 

\newtheorem{lemma}{Lemma}

\usepackage{ifthen}

\newcommand{\mbold}{1}

\ifthenelse{\mbold=1}{ 
\newcommand{\mathbold}{\mathbf} 
\newcommand{\bmbold}{\bm}}
{\newcommand{\mathbold}{} 
\newcommand{\bmbold}{}
}

\title{The inverse Kalman filter}
\author{Xinyi Fang and Mengyang Gu\footnote{The authors contribute equally. Correspondence should be addressed to Mengyang Gu (\href{mailto:mengyang@pstat.ucsb.edu}{mengyang@pstat.ucsb.edu})}}
\date{}

\begin{document}

\maketitle

\begin{abstract}
We introduce the inverse Kalman filter, which enables exact matrix-vector multiplication between a covariance matrix from a dynamic linear model and any real-valued vector with linear computational cost. 
We integrate the inverse Kalman filter with the conjugate gradient algorithm,  which substantially accelerates the computation of matrix inversion for a general form of covariance matrix, where other approximation approaches may not be directly applicable.  
We demonstrate the scalability and efficiency of the proposed approach through applications in nonparametric estimation of particle interaction functions, 
using both simulations and cell trajectories from microscopy data. 
\end{abstract}

\section{Introduction} 
\label{sec:intro}

\begin{figure}[t] 
\centering
\includegraphics[width=11cm]{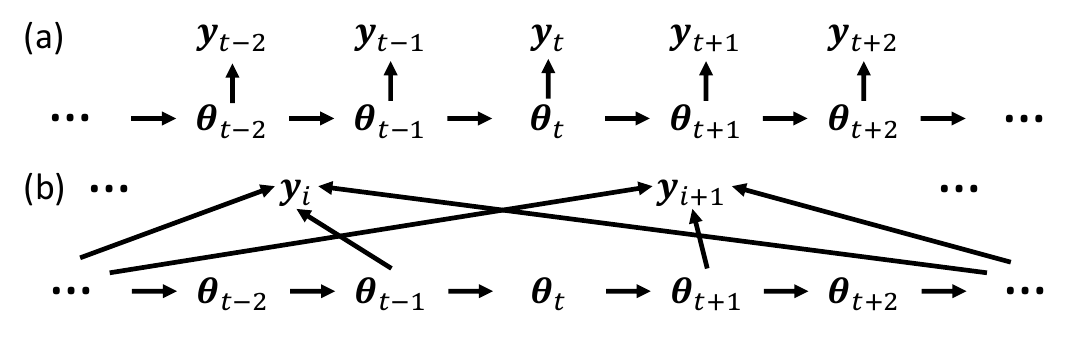}
\caption{The dependent structure of the (a) dynamic linear model and (b) particle dynamics. 
   } 
\label{fig:generalized_latent_factors}
\end{figure}

Dynamic linear models (DLMs) or linear state space models are ubiquitously used in modeling temporally correlated data \cite{West1997,durbin2012time}.  Each observation $y_t \in \mathbb R$  in DLMs 
is associated with a latent state vector $\bmbold \theta_t \in \mathbb R^q$, 
defined as   
\begin{align}
y_t&= \mathbold  F_{t}    \bmbold  \theta_t +   v_{t}, \quad  v_{t} \sim \mathcal{N}(0,   V_{t}),
\label{equ:DLM_y} \\
\bmbold  \theta_t&=   \mathbold  G_{t}   \bmbold  \theta_{t-1} +  \mathbold  w_{t},  \quad \mathbold  w_{t} \sim \mathcal{MN}( \mathbold  0,  \mathbold  W_{t}),
\label{equ:DLM_theta}
\end{align}
where $  \mathbold  F_{t}$ and $  \mathbold  G_{t}$ are matrices of dimensions $1\times q$ and $q \times q$, respectively, $\mathbold  W_{t}$ is a $q\times q$ covariance matrix for $t=2,\dots,N$ and the initial state vector follows a multivariate normal distribution $\bmbold  \theta_1\sim \mathcal{MN}(  \mathbold  b_{1},   \mathbold  W_{1}) $ with $\mathbold  b_{1} = \mathbold  0$ assumed herein. The dependent structure of the DLM is illustrated in Fig. \ref{fig:generalized_latent_factors}(a).

DLMs are a large class of models that include many widely used processes, such as autoregressive and moving average processes  \cite{petris2009dynamic,prado2010time}. Some Gaussian processes with commonly used covariance functions, including the Mat{\'e}rn covariance with a half-integer roughness parameter \cite{handcock1993bayesian}, can also be represented as DLMs \cite{whittle1954stationary,hartikainen2010kalman}, with closed-form expressions for $\mathbold  F_{t}$,  $\mathbold  G_{t}$ and $\mathbold  W_{t}$. This connection, summarized in the Supplementary Material, enables differentiable Gaussian processes to 
be used  for 
scalably estimating smooth latent functions from noisy observations.

The Kalman filter (KF) provides a fast approach for estimating latent states and computing likelihood functions in DLMs, which scales linearly with the number of observations 
\cite{kalman1960new}, as reviewed in Appendix A. 
In particular, the KF 
enables efficient computation of  $\mathbold L^{-1}\mathbold u$ for any $N$-dimensional vector $\mathbold u$ in $\mathcal O(q^3N)$ operations, where $\mathbold L$ is the Cholesky factor of the covariance matrix $\bmbold\Sigma=\mbox{cov}[\mathbold y_{1:N}]=\mathbold L\mathbold L^T$, summarized in Lemma \ref{lemma:L_inv_z} in the Appendix.

Computational challenges remain for high-dimensional state space models and scenarios where the KF cannot be directly applied, such as the dependent structure shown in Fig. \ref{fig:generalized_latent_factors}(b), where each observation $y_i$ can be associated with multiple latent states. This interaction structure, introduced in Section \ref{sec:application_particle}, is common in 
physical models, including molecular dynamics simulation \cite{rapaport2004art}.   
A way to overcome these challenges is to efficiently compute $\bmbold\Sigma\mathbold u$ for any $N$-dimensional vector $\mathbold u$ and utilize optimization methods such as conjugate gradient algorithms for computing predictive distributions \cite{hestenes1952methods}. This strategy has been used to approximate the maximum likelihood estimator of parameters in Gaussian processes \cite{stein2013stochastic,majumder2022kryging}. Yet, each conjugate gradient iteration requires matrix-vector multiplication, which involves $\mathcal O(N^2)$ operations and storage, making it prohibitive for large $N$.

To address this issue, we introduce the \textit{inverse Kalman filter} (IKF), which computes $\mathbold L\mathbold u$ and $\mathbold L^T\mathbold u$ 
with $\mathcal O(q^3N)$ operations without approximation, where  $\mathbold L$ is the Cholesky factor of any $N \times N$ DLM-induced covariance $\bmbold  \Sigma$ and $\mathbold u$ is an $N$-dimensional vector.  
The complexity is significantly smaller than $\mathcal O(N^2)$ in direct computation. 
The dimension of the latent states $q$ does 
not exceed 3 in our applications, and this choice includes many commonly used models such as twice differentiable Gaussian processes with Mat{\'e}rn covariance in (\ref{equ:matern_5_2}), often assumed as default in Gaussian process models.

The IKF algorithm can be extended to accelerate matrix multiplication and inversion, a key computational bottleneck in practice. We integrate the IKF into the conjugate gradient algorithm to scalably compute predictive distributions of observations with a general covariance structure:  
\begin{equation}
\bmbold \Sigma_y= \sum^{J}_{j=1}   \mathbold  A_j   \bmbold  \Sigma_j   \mathbold  A^T_j +\bmbold \Lambda,
\label{equ:Sigma_y}
\end{equation}
where $\bmbold  \Sigma_j$ is a DLM-induced covariance matrix, $\mathbold  A_j$ is sparse, and $\bmbold \Lambda$ is a diagonal matrix. This structure appears 
in nonparametric estimation of particle interactions, which will be introduced in Section \ref{sec:application_particle}. Both real and simulated studies involve numerous particles over a long time period, where conventional approximation methods \cite{vecchia1988estimation,katzfuss2022scaled,lindgren2011explicit,datta2016hierarchical,finley2022spnngp, gramacy2015local} are not applicable. This covariance structure also appears in other applications, including varying coefficient models \cite{hastie1993varying} 
and estimating incomplete lattices of correlated data \cite{stein2012interpolation}. 
Though we focus on nonparametric estimation of particle interaction functions in this work, we also apply our approach to the latter in Section S5 and provide numerical comparisons with various approximation methods in Section S9 of the Supplementary Material.

\section{Inverse Kalman filter}
\label{sec:IKF}

In this section, we introduce an exact algorithm for computing $\bmbold\Sigma \mathbold{u}$ with $\mathcal{O}(q^3 N)$, where $\mathbold{u}$ is an $N$-dimensional real-valued vector and $\bmbold\Sigma=\mbox{cov}[\mathbold y_{1:N}]$ is an $N \times N$ covariance matrix of observations induced by a DLM with $q$ latent states in  (\ref{equ:DLM_y})-(\ref{equ:DLM_theta}). This algorithm is applicable to any DLM specified in (\ref{equ:DLM_y})-(\ref{equ:DLM_theta}).  
Define $\bmbold \Sigma= \mathbold L \mathbold {L}^T$, where the Cholesky factor $\mathbold L$ does not need to be explicitly computed in our approach. 
We develop fast algorithms to compute $\mathbold {\tilde x}= \mathbold  L^T \mathbold {u}$ and $\mathbold  {x}= \mathbold  L   \mathbold  {\tilde x}$ by Lemma \ref{lemma:L_t_tilde_u} and Lemma \ref{lemma:L_tilde_x} below, respectively, each with $\mathcal O(q^3N)$ operations. Detailed proofs of these lemmas are available in the Supplementary Material. 
In the following, $u_t$, $x_t$, $\tilde x_t$ denote the $t$th elements of the vectors $  \mathbold  {u}$, $  \mathbold  x$, and $  \mathbold  {\tilde x}$, respectively, and $L_{t',t}$ denotes the $(t',t)$th element of $\mathbold{L}$ for $t$ and $t'=1,2,\dots,N$.

\begin{lemma}[Compute $\tilde{  \mathbold  x}=   \mathbold {L}^T {  \mathbold  u}$ with linear computational cost] 
For any $N$-dimensional vector ${  \mathbold  u}$, 
let $\tilde x_{N}={Q_{N}^{1/2}} u_{N}$, $\tilde x_{N-1}={Q_{N-1}^{1/2}}( \ell_{N,N-1} u_{N}+u_{N-1})$, and $\mathbold  g_{N-1}=    \mathbold  F_{N}   \mathbold  G_{N} u_{N}$. 
For $t=N-2,\dots,1$, iteratively compute $\tilde x_t$ using
\begin{align}
\tilde x_t &= {Q_t^{\frac{1}{2}}}\left(\tilde {\ell}_{t+1,t} +  \ell_{t+1,t} u_{t+1}+  u_t\right), \label{equ:tilde_x_t} \\ 
{  \mathbold  g}_t &=   \mathbold  g_{t+1} {  \mathbold  G}_{t+1}+  \mathbold  F_{t+1}   \mathbold  G_{t+1} u_{t+1}, \label{equ:g_t} 
\end{align}
with $\ell_{t+1,t}=  \mathbold  F_{t+1}  \mathbold  G_{t+1}   \mathbold  K_t$,  
$\tilde {\ell}_{t+1,t}={  \mathbold  g}_{t+1}  \mathbold  G_{t+1}   \mathbold  K_t$, 
and the Kalman gain 
\begin{align}
  \mathbold  K_t=   \mathbold  B_t    \mathbold  F^T_t  Q^{-1}_t,  
\label{equ:K_t}
\end{align}
where $\mathbold  B_t=\mbox{cov}[   \bmbold  \theta_t \mid {  \mathbold   y}_{1:t-1}]$  and $Q_t=\mbox{var}[  y_t \mid {  \mathbold  y}_{1:t-1}]$ are respectively defined in the Kalman filter in (\ref{equ:KF1}) and (\ref{equ:KF2}) in Appendix A. Then $\tilde{  \mathbold  x}=   \mathbold {L}^T {  \mathbold  u}$. 
 \label{lemma:L_t_tilde_u}
\end{lemma}
	 
\begin{lemma}[Compute $\mathbold  x=   \mathbold   {L} \tilde{  \mathbold  x}$ with linear computational cost]
For any $N$-dimensional vector $\tilde{  \mathbold  x}$, 
let $x_1=  \mathbold  F_1  {  \mathbold  b}_1+{Q_1}^{1/2} \tilde x_1$ and $\tilde{  \mathbold  m}_1= {  \mathbold  b}_1+  \mathbold  K_1 (x_1-  \mathbold  F_1  {  \mathbold  b}_1)$.   For $t=2,\dots,N$,  iteratively compute   $x_t$ using 
\begin{align}
{  \mathbold  b}_t&=  \mathbold  G_t \tilde{  \mathbold  m}_{t-1}, \label{equ:b_t}\\
x_t&=  \mathbold  F_t  {  \mathbold  b}_t+{Q_t^{\frac{1}{2}}} \tilde x_t, \label{equ:x_t} \\
\tilde{  \mathbold  m}_t&= {  \mathbold  b}_t+  \mathbold  K_t (x_t-  \mathbold  F_t  {  \mathbold  b}_t),  \label{equ:tilde_m_t}
\end{align}
where $  \mathbold  K_t$  and $Q_t$ are respectively defined in (\ref{equ:K_t}) and (\ref{equ:KF2}) in the Appendix. Then $\mathbold  x=   \mathbold   {L} \tilde{  \mathbold  x}$. 
      
      \label{lemma:L_tilde_x}
\end{lemma}

The algorithm in Lemma \ref{lemma:L_t_tilde_u} is unconventional and nontrivial to derive. In contrast, Lemma \ref{lemma:L_tilde_x} has a direct connection to the Kalman filter. Specifically, the one-step-ahead predictive distribution in step (ii) of the Kalman filter, given in Appendix A, enables computing $\mathbold  {L}^{-1}   \mathbold x$ for any vector $\mathbold  x$ (Lemma \ref{lemma:L_inv_z}). Lemma \ref{lemma:L_tilde_x} essentially reverses this operation, computing $  \mathbold  {L}   \mathbold  {\tilde x}$ for any vector $\mathbold {\tilde x}$, leading to the term \textit{Inverse Kalman filter} (IKF).

The IKF algorithm, outlined in Algorithm \ref{alg:robust_IKF}, sequentially applies Lemmas \ref{lemma:L_t_tilde_u} and \ref{lemma:L_tilde_x} to reduce the computational cost and storage for $\bmbold  \Sigma {\mathbold  u}$ from $\mathcal O(N^2)$ to $\mathcal O(q^3 N)$ without approximation.  Note that the $\mathcal O(q^3N)$ is primarily due to the matrix-matrix multiplication required to obtain $\mathbf B_t$ (Eq.~(\ref{equ:KF1})) in the initial KF. Once these quantities are pre-computed, the core iterations in Lemmas \ref{lemma:L_t_tilde_u}-\ref{lemma:L_tilde_x} only involve matrix-vector multiplications, which require $\mathcal O(q^2N)$.
The IKF applies to all DLMs, including Gaussian processes that admit equivalent DLM representation, 
such as Gaussian processes with Mat{\'e}rn kernels where roughness parameters $0.5$ and $2.5$ correspond to $q=1$ and $q=3$, respectively. Details are provided in the Supplementary Material.

\vspace{.5em}


\begin{algorithm}[t]
\caption{
The IKF for computing $  \bmbold  \Sigma   \mathbold {u}$} \label{alg:robust_IKF}
\begin{algorithmic}[1]
\Input{$  \mathbold  F_t,  \mathbold  G_t,   \mathbold  W_t, t=1, \dots, N$,  an $N$-dimensional vector ${  \mathbold  u}$.}
\State Use Kalman filter from Lemma \ref{lemma:KF} to compute $  \mathbold  B_t$, $Q_t$, and $  \mathbold  K_t$ for $t=1,\dots,N$.
\State Use Lemma \ref{lemma:L_t_tilde_u} to compute  $\tilde{  \mathbold  x}=   \mathbold   {L}^T {  \mathbold  u}$.
\State Use Lemma \ref{lemma:L_tilde_x} to compute  ${  \mathbold  x}=   \mathbold  {L} \tilde{  \mathbold  x}$. 
\Output{ ${  \mathbold  x}$. 
}
\end{algorithmic}
\end{algorithm}

\begin{figure}[t] 
\centering
\includegraphics[width=16cm]{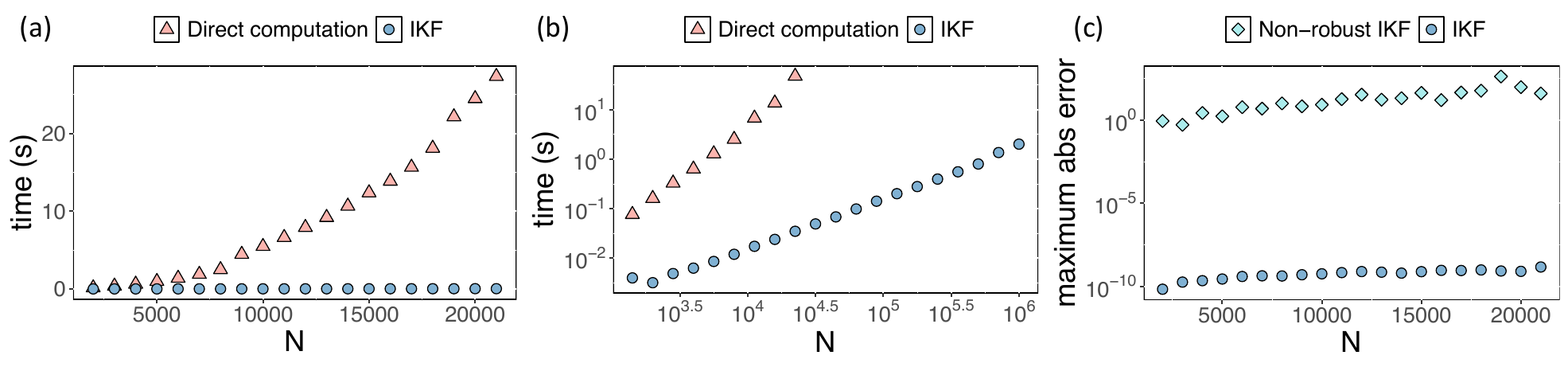}
\caption{Panels (a) and (b) compare the IKF (blue dots) with direct computation (red triangles) for calculating $  \bmbold  \Sigma   \mathbold {u}$ in the original time scale and logarithmic scale (base 10), respectively. Panel (c) shows the predictive error comparison between the non-robust IKF (light blue squares) and the robust IKF (blue dots).  }
\label{fig:comparison_time_error}
\end{figure}

While Algorithm \ref{alg:robust_IKF} can be applied to a DLM regardless of the noise variance, we do not recommend directly computing $\bmbold\Sigma\mathbold u$ in scenarios when the noise variance $V_t$ is zero or close to zero in (\ref{equ:DLM_y}). This is because, when $V_t=0$, $Q_t$ in (\ref{equ:KF2}), the variance of the one-step-ahead predictive distribution, could be extremely small, which leads to large numerical errors in  (\ref{equ:KF3}) due to the unstable inversion of $Q_t$. To ensure robustness, we suggest computing $  \mathbold  x=   (  \bmbold  \Sigma+ V\mathbold  I_{N}) {  \mathbold u}$ with  a positive scalar $V$ and outputting $  \mathbold  x- V {  \mathbold  u}$ as the result of $  \bmbold  \Sigma {  \mathbold  u}$.  Here, $\bmbold  \Sigma+ V\mathbold  I_{N}$ can be interpreted as the covariance matrix of a  DLM with noise variance $V_t = V$. Essentially,  we introduce artificial noise to ensure that $Q_t$ computed in the KF is at least $V$ for numerical stability. The result remains exact and independent of the choice of $V$.

We compare the IKF approach with the direct matrix-vector multiplication of $\bmbold  \Sigma   \mathbold {u}$ in noise-free scenarios in Fig. \ref{fig:comparison_time_error}. 
The inputs of $\bmbold \Sigma$ are uniformly sampled from $[0,1]$ and the Mat{\'e}rn covariance with unit variance,  roughness parameter $2.5$, and range parameter $\gamma=0.1$ \cite{handcock1993bayesian} is used. 
Panels (a) and (b) show that the IKF significantly reduces computational costs compared to direct computation.
The IKF achieves covariance matrix-vector multiplication for an output vector of  $10^6$ dimensions in about 2 seconds on a 
desktop, making it highly scalable to be applied in iterative algorithms like the conjugate gradient algorithm \cite{hestenes1952methods} and the randomized log-determinant approximation \cite{saibaba2017randomized}. 
Fig. \ref{fig:comparison_time_error}(c) compares the maximum absolute error between the robust IKF with $V=0.1$ and the non-robust IKF with $V=0$.
The non-robust IKF exhibits large numerical errors due to instability when $Q_t$ approaches zero, while the robust IKF remains stable by ensuring that $Q_t$ is no smaller than $V$, even with near-singular covariance matrices $\bmbold{\Sigma}$.

Lemmas \ref{lemma:L_t_tilde_u} and \ref{lemma:L_tilde_x} facilitate the computation of each element in $\mathbold{L}$ using DLM parameters and enable the linear computation of $(\mathbold {L}^T)^{-1}\tilde{\mathbold x}$. 
These results are respectively summarized in Lemmas \ref{lemma:chol_KF} and \ref{lemma:L_t_inv_tilde_x} below, with proofs provided in the Supplementary Material. 

\begin{lemma}[Cholesky factor from the inverse  Kalman filter]
Each entry $(t',t)$ in the lower triangular matrix $\mathbold  {L}$ with $t' \geq t$ has the form
\begin{align}
L_{t',t}={Q_t^{\frac{1}{2}}} \ell_{t',t}, 
\label{equ:L_k_t}
\end{align}
where $ \ell_{t',t'}=1$, and, for $t'>t$, $  \ell_{t',t}$ is defined as  
\begin{align}
 \ell_{t',t}=  \mathbold  F_{t'} \left(\prod^{t'}_{l=t+1}  \mathbold  G_l\right)  \mathbold  K_t, 
 \label{equ:u}
\end{align}
with $\prod^{t'}_{l=t+1}  \mathbold  G_l=  \mathbold  G_{t'}  \mathbold  G_{t'-1}\dots  \mathbold  G_{t+1}$,  $  \mathbold  K_t$  and $Q_t$ respectively defined 
 in  (\ref{equ:K_t}) and  (\ref{equ:KF2}). 
        \label{lemma:chol_KF}
\end{lemma}

\begin{lemma}[Compute ${ \mathbold  u}=(\mathbold {L}^T)^{-1}\tilde{ \mathbold  x}$ with linear computational cost] For any $\tilde {\mathbold x}$, 
let $u_{N}={Q_{N}^{-1/2}}\tilde x_{N} $,   
and $u_{N-1}=Q_{N-1}^{-1/2}\tilde x_{N-1}-\ell_{N,N-1} u_{N}$.  For $t=N-2,\dots,1$, iteratively compute $u_t$ using
\begin{align}
u_t &=Q_t^{-\frac{1}{2}}\tilde x_t-{\tilde \ell}_{t+1,t}-  \ell_{t+1,t} u_{t+1}, 
\label{equ:u_t}
\end{align}
with ${\tilde \ell}_{t+1,t}$ and $\ell_{t+1,t}$ defined in Lemma \ref{lemma:L_t_tilde_u}. 
Then ${ \mathbold  u}=(\mathbold {L}^T)^{-1}\tilde{ \mathbold  x}$. 
\label{lemma:L_t_inv_tilde_x}
\end{lemma}

\section{Nonparametric estimation of particle interaction functions}
\label{sec:application_particle}
The IKF algorithm is motivated by the problem of estimating interactions between particles, which is crucial to understanding complex behaviors of molecules and active matter, such as migrating cells and flocking birds. Minimal physical models, such as the Vicsek model \cite{vicsek1995novel} 
and its extensions \cite{chate2008modeling,ginelli2010large}, provide a framework for explaining the collective motion of active matter.

Consider a physical model encompassing multiple types of latent interactions. {Let $\mathbold y_i$ be a $D_y$-dimensional output vector 
corresponding to particle $i'$ at time $\tau$,}   
which is influenced by 
$J$ distinct interaction types.  
For the $j$th type of interaction, the $i$th observational vector interacts with a subset $p_{i,j}$ of particles rather than with all particles, typically those within a radius $r_j$. This relationship is expressed as a latent factor model:
\begin{equation}
  \mathbold  y_i= \sum^J_{j=1} \sum^{p_{i,j}}_{k=1}    \mathbold  a_{i,j,k}  z_{j}(d_{i,j,k})+    \bmbold  \epsilon_i, 
\label{equ:dynamic_factors}
\end{equation}
with  $\bmbold  \epsilon_i\sim \mathcal{MN}(0, \sigma^2_0I_{D_y})$ denoting a Gaussian noise vector. The term $\mathbold  a_{i,j,k}$ is a $D_y$-dimensional \textit{known} factor loading that links the $i$th output to the $k$th neighbor in the $j$th interaction, with the \textit{unknown}  interaction function $z_j(\cdot)$ evaluated at a scalar-valued input $d_{i,j,k}$, such as the distance between particles $i$ and $k$, for $k=1,\dots,p_{i,j}$, $i=1,\dots,n$, and $j=1,\dots,J$. For a dataset of $n_p$ particles over $n_{\tau}$ time points, the total number of observations is $\tilde N=nD_y=n_p n_{\tau}D_y$.

{An illustrative example of this framework is an unnormalized Vicsek model. Unlike the original Vicsek model \cite{vicsek1995novel}, which enforces a constant speed for all particles, we allow velocity magnitudes to vary over time. Time-varying velocities are more sensible in many real-world scenarios, including cellular movements studied in Section \ref{sec:cell}. When the system approaches a high degree of confluence, the average velocity of the cells typically decreases because of cell proliferation.  
Assume that the 2-dimensional velocity of the $i'$th particle at time $\tau$, $\mathbold v_{i'}(\tau)=(v_{i',1}(\tau),v_{i',2}(\tau))^T$, is updated by averaging the velocity of its neighbors from the previous time step. Specifically, for each direction $l=1,2$,
\begin{align}
    v_{i',l}(\tau)=\frac{1}{p_{i'}(\tau-1)}\sum_{k \in ne_{i'}(\tau-1) } v_{k,l}(\tau-1) + \epsilon_{i',l}(\tau),
    \label{equ:unnormalized_Vicsek_v}
\end{align}
where $\epsilon_{i'}(\tau)$ is a zero-mean Gaussian noise with variance $\sigma^2_{0}$. The set of neighbors $ne_{i'}(\tau-1)$ includes particles within a radius of $r$ from particle $i'$ at time $\tau-1$, i.e., $ne_{i'}(\tau-1)=\{k: ||  \mathbold  s_{i'}(\tau-1)-  \mathbold  s_{k}(\tau-1)||<r\}$, where $  \mathbold  s_{i'}(\tau-1)$ and $  \mathbold  s_{k}(\tau-1)$ are 2-dimensional position vectors of particle $i'$ and its neighbor $k$, respectively, and $p_{i'}(\tau-1)$ denotes the total number of neighbors of the $i'$th particle, including itself, at time $\tau-1$. The illustration of this model is shown in Fig.~S1(b) 
in the Supplementary Material.

The unnormalized Vicsek model in (\ref{equ:unnormalized_Vicsek_v}) is a special case of the general framework in  (\ref{equ:dynamic_factors}) with scalar output $\mathbold y_{i}=v_{i',l}(\tau)$, i.e. $D_y=1$, and the index of the $i$th observation $i=2(\tau-1)n_p+2(i'-1)+l$ is a function of time $\tau$ and particle $i'$. 
This model contains a single interaction, i.e. $J=1$, with a linear interaction function $z(d)=d$, where $d$ is the velocity component of the neighboring particle, and the factor loading being $1/p_{i'}(\tau-1)$. 
}

Numerous other physical models of self-propelled particles can also be formulated as in (\ref{equ:dynamic_factors}) with a parametric form of a particle interaction function \cite{chate2008modeling}. Nonparametric estimation of particle interactions is preferred when the physical mechanism is unknown \cite{katz2011inferring,lu2019nonparametric}, but the high computational cost limits its applicability to systems with large numbers of particles over long time periods. 

In this work, we model each latent interaction function $z_j(\cdot)$ nonparametrically using a Gaussian process. By utilizing an equivalent DLM representation, we can leverage the 
proposed IKF algorithm to significantly expedite computation. This includes commonly used kernels such as the Mat{\'e}rn kernel with a half-integer roughness parameter 
\cite{handcock1993bayesian}, often used as the default setting for Gaussian process models to predict nonlinear functions, as the smoothness of the process can be controlled by its roughness parameters \cite{Gu2018robustness}. 

For each interaction $j$, we form the input vector $\mathbold  d^{(u)}_j=[d^{(u)}_{1,j},\dots,d^{(u)}_{N_j,j}]^T$, where superscript $(u)$ indicates that the input entries are unordered, $d^{(u)}_{t,j}=d_{i,j,k}$ with $t=\sum^{i-1}_{i'=1} p_{i',j}+k$ for any tuple $(i,j,k)$, $k = 1, \dots, p_{i,j}$, $i=1,\dots,n$, $j=1,\dots,J$, and $N_j=\sum^{n}_{i=1}p_{i,j}$. 
The marginal distribution of the $j$th factor follows $\mathbold  z_j^{(u)}=[z_j(d_{1,j}^{(u)}),\dots,z_j(d_{N_j,j}^{(u)})]^T \sim \mathcal{MN} (  \mathbold  0, \,   \bmbold  \Sigma_j^{(u)})$, where the $(t,t')$th entry of the $N_j\times N_j$ covariance matrix $\bmbold  \Sigma_j^{(u)}$ is $\mbox{cov}[z_j(d_{t,j}^{(u)}),z_j(d_{t',j}^{(u)})] = \sigma^2_jc_j(d_{t,j}^{(u)},d_{t',j}^{(u)}) = \sigma^2_jc_j(d)$, with $d=|d_{t,j}^{(u)}-d_{t',j}^{(u)}|$, and $\sigma^2_j$ and $c_j(\cdot)$ the variance parameter and correlation function, respectively. 
We employ the Mat{\'e}rn correlation with roughness parameters $\nu_j = 0.5$ and $\nu_j = 2.5$. For $\nu_j=0.5$, the Mat{\'e}rn correlation is  the exponential correlation $c_j(d)=\exp(-d/\gamma_j)$, while for $\nu_j=2.5$, the Mat{\'e}rn correlation follows
\begin{equation}
 c_j(d)= \left(1+\frac{{5}^{1/2}d}{\gamma_j}+\frac{5d^2}{3\gamma_j^2}\right)\exp\left(-\frac{{5}^{1/2}d}{\gamma_j}\right) \,, 
\label{equ:matern_5_2}
\end{equation}
where $\gamma_j$ denotes the range parameter.

By integrating out latent factor processes $\mathbold  z_j$, 
the marginal distribution of the observational vector $\mathbold  y=(\mathbold  y^T_1,\dots,  \mathbold  y^T_n)^T$, with dimension $\tilde N=nD_y$,  follows a multivariate normal distribution: 
\begin{align}
	 \left(  \mathbold  y \mid   \bmbold  \Sigma^{(u)}_j, \sigma^2_0 \right) \sim \mathcal{MN}\left(  \mathbold  0, \sum^{J}_{j=1}   \mathbold  A_j   \bmbold  \Sigma^{(u)}_j   \mathbold  A^T_j +\sigma^2_0   \mathbold  I_{\tilde N} \right).
\label{equ:marginal_lik}
\end{align}
Here, $\mathbold  A_j$ is a sparse $\tilde N \times N_j$ block diagonal matrix, such that the $i$th diagonal block is a $D_y\times p_{i,j}$ matrix $\mathbold  A_{i,j}=[\mathbold  a_{i,j,1},\dots, \mathbold  a_{i,j,p_{i,j}}]$ for $i=1,\dots,n$. 
The posterior predictive distribution of the latent variable $z_j(d^*)$ for any given input $d^*$ is also a normal distribution: 
\begin{equation}
\left(z_j(d^*) \mid   \mathbold  y,   \sigma^2_0, \bmbold \sigma^2,  \bmbold { \gamma},  \mathbold   r  \right) \sim \mathcal N(\hat z_j(d^*), \sigma^2_jc^*_j(d^*)),
\label{equ:post_z_j}
\end{equation}
with the predictive mean and variance given by  
\begin{align}
\hat z_j(d^*)&=   \bmbold  \Sigma^{(u)}_j(d^{*})^T  \mathbold  A^T_j   \bmbold  \Sigma_y^{-1}   \mathbold  y, \label{equ:hat_z_j}\\
\sigma^2_jc^*_j(d^*)&= \sigma^2_jc_j(d^*,d^*)-   \bmbold  \Sigma^{(u)}_j(d^*)^T   \mathbold  A^T_j  \bmbold   \Sigma_y^{-1}   \mathbold  A_j   \bmbold  \Sigma^{(u)}_j(d^*).
\label{equ:K_star_j}
\end{align}
Here, $  \bmbold  \Sigma_y=\sum^{J}_{j=1}   \mathbold  A_j  \bmbold  \Sigma^{(u)}_j   \mathbold  A^T_j +\sigma^2_0   \mathbold  I_{\tilde N}$,  $  \bmbold  \Sigma^{(u)}_j(d^*)=\sigma^2_j[c_j(d^{(u)}_{1,j},d^*),\dots,c_j(d^{(u)}_{ N_j,j}, d^*)]^T$  and $\mathbold r$ represents additional system parameters. 

The main computational challenge in calculating the predictive distribution in (\ref{equ:post_z_j}) lies in inverting the covariance matrix $\bmbold  \Sigma_y$, where $\tilde N$ and $N_j$ range between $10^5$ and $10^6$ in our applications. Though various Gaussian process approximation methods have been proposed  \cite{vecchia1988estimation,snelson2006sparse,furrer2006covariance,cressie2008fixed,lindgren2011explicit,gramacy2015local,datta2016hierarchical,katzfuss2021general}, they typically approximate the parametric covariance $\bmbold  \Sigma_j^{(u)}$ rather than $\bmbold  \Sigma_y$, and thus they are not directly applicable  for computing the distribution in (\ref{equ:post_z_j}).

To address this, we employ the conjugate gradient (CG) algorithm \cite{hestenes1952methods} to accelerate the computation of the predictive distribution. Each conjugate gradient iteration needs to compute $\bmbold  \Sigma_y \mathbold u$ for an $\tilde N$-dimensional vector ${\mathbold u}$.  To employ the IKF, we first  rearrange the input vector $\mathbold  d^{(u)}_j$ into a non-decreasing input sequence 
$\mathbold d_j = [d_{1,j},\dots,d_{N_j,j}]^T$. 
Denote the covariance $\mbox{var}[\mathbold z_j]=\bmbold \Sigma_j$ with $\mathbold z_j=[ z_j(d_{1,j}),\dots, z_j(d_{N_j,j})]^T$. 
The covariance of the observations can be written as a weighted sum of $\bmbold \Sigma_j$ by introducing a permutation matrix for each interaction: $\bmbold \Sigma_y=\sum^{J}_{j=1}    \mathbold  A_{\pi, j} \bmbold  \Sigma_j    \mathbold  A^T_{\pi,j} +\sigma^2_0   \mathbold  I_{N}$, where $ \mathbold  A_{\pi, j}  =\mathbold  A_{j}  \mathbold P^T_j$ with  $\mathbold P_j$ a permutation matrix such that $\bmbold  \Sigma_j=\mathbold P_j \bmbold  \Sigma^{(u)}_j    \mathbold P^T_j $. After this reordering, the computation of $\bmbold  \Sigma_y \mathbold u$  can be broken into  four steps: 
\begin{enumerate}[label=(\arabic*), itemsep=0pt, parsep=0pt, topsep=3pt]
    \item $\mathbold u_j = \mathbold A_{\pi,j}^T \mathbold u$, 
    \item $\mathbold  x_j= \bmbold  \Sigma_j \mathbold {u}_j$,
    \item $ \mathbold  {\hat u}_j= \mathbold  A_{\pi,j}  \mathbold  x_j$,
    \item $\sum^J_{j=1}  \mathbold {\hat u}_j+\sigma^2_0 \mathbold  u$. 
\end{enumerate}
Here, $\mathbold A_j$ is a sparse matrix with $N_jD_y$ non-zero entries. 
The IKF algorithm is used in step (2) to accelerate the most expensive computation, with all computations performed directly using terms in the KF algorithm without explicitly constructing the covariance matrix.

We refer to the resulting approach as the IKF-CG algorithm. This approach reduces the computational cost for computing the posterior distribution from $\mathcal O(\tilde N^3)$ operations to pseudolinear order with respect to $N_j$, as shown in Table S1 in the Supplementary Material. 
Furthermore, the IKF-CG algorithm can accelerate the parameter estimation via both cross-validation and maximum likelihood estimation, with the latter requiring an additional approximation of the log-determinant \cite{saibaba2017randomized}. Details of the conjugate gradient algorithm, the computation of the predictive distribution, parameter estimation procedures, and computational complexity are discussed in Sections S3, S4, S6, and S7 of the Supplementary Material, respectively.

\section{Numerical results for particle interaction function estimation}
\label{sec:num_results_particle}
\subsection{Evaluation criteria}
We conduct simulation studies in Sections \ref{sec:Vicsek}-\ref{sec:modify_Vicsek} and analyze real cell trajectory data in Section \ref{sec:cell} to estimate particle interaction functions in physical models. The code and data to reproduce all numerical results are available at \url{https://github.com/UncertaintyQuantification/IKF}. We have implemented the main functions of the IKF and IKF-CG algorithms for estimating particle interaction functions in the {\tt FastGaSP} package on CRAN \cite{gu2025fastgasppackage}. 
The observations in simulation are generated at equally spaced time frames $\tau=1,\dots, n_{\tau}$ with interval $h=0.1$, though the proposed approach is applicable to both equally and unequally spaced time frames. 
For simplicity, the number of particles $n_{p}$ is assumed constant across all time frames during simulations.

For each of the $J$ latent factors, predictive performance is assessed using the normalized root mean squared error (NRMSE$_j$), the average length of the $95\%$ posterior credible interval ($L_j(95\%)$), and the proportion of interaction function values covered within the 95\% posterior credible interval ($P_j(95\%)$), based on $n^*$ test inputs $  \mathbold  d^*_j=(d^*_{1,j},\dots,d^*_{n^*,j})^T$:
\begin{align}
\mbox{NRMSE}_j&=\left(\frac{{\sum^{n^*}_{i=1} (\hat z_j(d^*_{i,j})-  z_j(d^*_{i,j}))^2}}{\sum^{n^*}_{i=1} (\bar z_j-  z_j(d^*_{i,j}))^2}\right)^{1/2},\label{equ:nrmse}\\
L_{j}(95\%)&=\frac{1}{n^*}\sum^{n^*}_{i=1}\mbox{length}\left\{ CI_{i,j}(95\%) \right\}, \label{equ:length}\\
P_j(95\%)&=\frac{1}{n^*}\sum^{n^*}_{i=1}1_{z_j(d^*_{i,j}) \in CI_{i,j}(95\%)} \label{equ:coverage}.
\end{align}
Here, $\hat z_j(d^*_{i,j})$ represents the predicted mean at $d^*_{i,j}$, $\bar z_j$ is the average of the $j$th interaction function, 
and $CI_{i,j}(95\%)$ denotes the $95\%$ posterior credible interval of $z_j(d^*_{i,j})$, for $j=1,\dots,J$. A desirable method should have a low NRMSE$_j$, small $L_{j}(95\%)$, and $P_j(95\%)$ close to $95\%$. 

To account for variability due to initial particle positions, velocities, and noise, each simulation scenario is repeated $E = 20$ times to compute the average performance metrics. 
Unless otherwise specified, parameters are estimated by cross-validation, with $80\%$ of the data used as a training set 
and the remaining $20\%$ as the validation set. 
All computations are performed on a macOS Mojave system with an 8-core Intel i9 processor running at 3.60 GHz and 32 GB of RAM.

\subsection{Unnormalized Vicsek model}
\label{sec:Vicsek}

\begin{figure}[t]
    \centering
    \includegraphics[width=16cm]{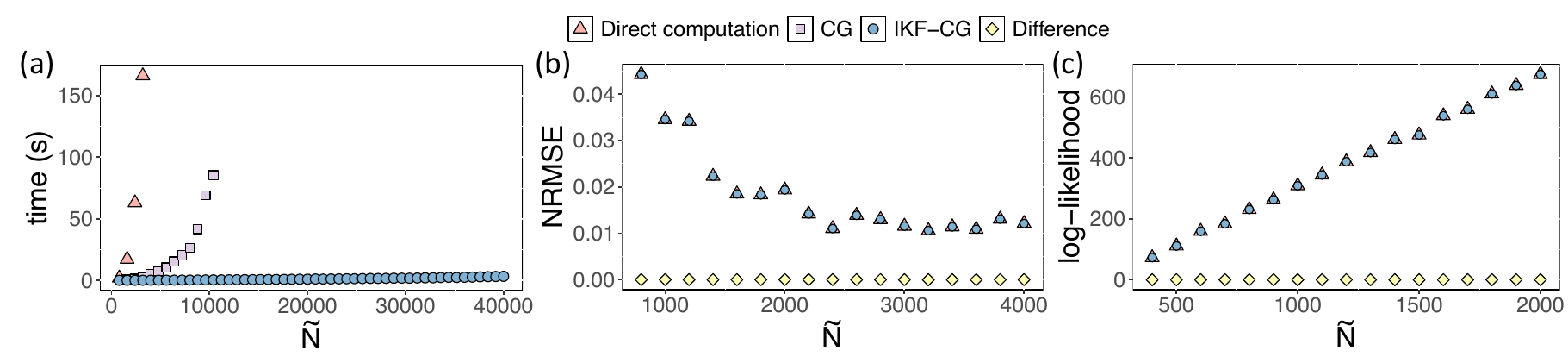}
    \caption{{(a) Computational time of calculating predictive means using the IKF-CG algorithm (blue dots), CG method (purple squares), and direct computation (red triangles) for the unnormalized Vicsek model with different numbers of observations. (b), (c) NRMSE of predictive means for the interaction function and log-likelihood computed via direct computation (red triangles) and the IKF-CG algorithm (blue dots), with differences (yellow diamonds) of orders $10^{-7}$ to $10^{-5}$ and $10^{-5}$ to $10^{-4}$, respectively.}}
\label{fig:time_approx_error_unormalized_Vicsek}
\end{figure}

{We first consider the unnormalized Vicsek model introduced in Section \ref{sec:application_particle}. For each particle $i'$  at time $\tau$, after obtaining its velocity using (\ref{equ:unnormalized_Vicsek_v}), its position is updated as:
\begin{align}
  \mathbold  s_{i'}(\tau)&=   \mathbold  s_{i'}(\tau-1)+ \mathbold v_{i'}(\tau)  h, \quad i'=1,\dots,n_{p},
\label{equ:Vicsek_s_k}
\end{align}
where $h$ is the time step size. Particles are initialized with velocity $[v\cos(\phi_{i'}(0)), v\sin(\phi_{i'}(0))]^T$, where $\phi_{i'}(0)$ is drawn from a uniform distribution on $[-\pi,\pi]$ and $v={2}^{1/2}/2\approx 0.71$. Initial particle positions $\mathbold s_{i'}(0)$ in (\ref{equ:Vicsek_s_k}) are uniformly sampled from $[0,{n_{p}}^{1/2}]\times [0,{n_{p}}^{1/2}]$ to keep consistent particle density across experiments. The goal is to estimate the interaction function nonparametrically, without assuming linearity. The interaction radius $r = 0.5$ is estimated alongside other parameters, with the estimation results detailed in the Supplementary Material.

\begin{figure}[tp]
\centering 
\includegraphics[width=15cm]{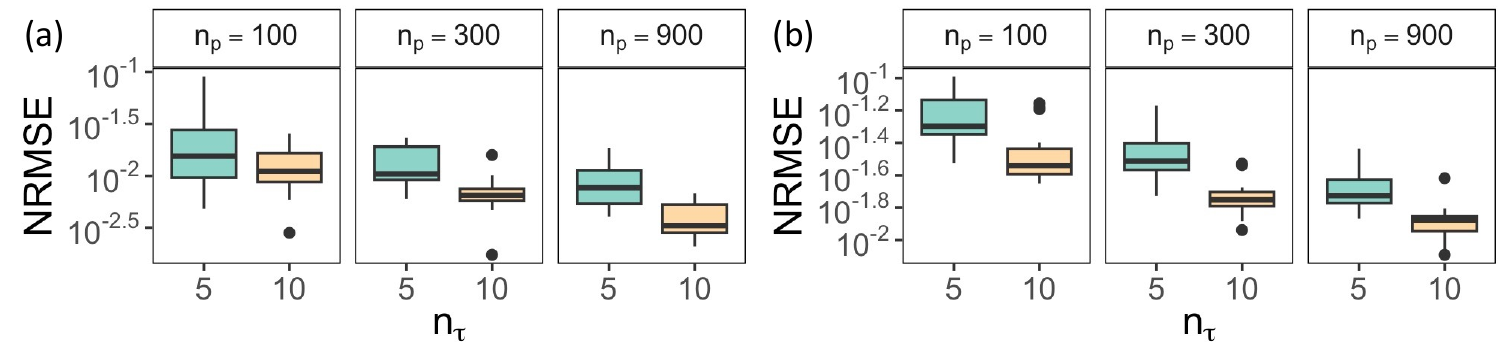}
\caption{{Boxplots of NRMSE in (\ref{equ:nrmse}) for estimating the latent interaction function in the unnormalized Vicsek model with $\sigma_0^2 = 0.1^2$ based on 20 experiments. Panel (a) uses the Mat{\'e}rn covariance with roughness parameter 2.5, and panel (b) uses the exponential covariance.} 
}
\label{fig:unnormalized_Vicsek_nrmse_0.1}
\end{figure}

We first compare the computational cost and accuracy of the IKF-CG, CG, and direct computation for calculating the predictive mean in  (\ref{equ:hat_z_j}) and the marginal likelihood in (\ref{equ:marginal_lik}) using the covariance in (\ref{equ:matern_5_2}). Simulations are conducted with $n_{p}=100$ particles and noise variance $\sigma^2_0=0.2^2$. 
Figure\ref{fig:time_approx_error_unormalized_Vicsek}(a) shows that the IKF-CG approach substantially outperforms both direct computation and the 
CG algorithm in computational time when predicting $n^*=100$ test inputs with varying time lengths $n_{\tau}$, ranging from 4 to 200. 
Figure \ref{fig:time_approx_error_unormalized_Vicsek}(b) compares the NRMSE of the predictive mean between the direct computation and IKF-CG method when the number of observations ranges from 400 to 2,000, corresponding to $n_{\tau}$ from 4 to 20. The two methods yield nearly identical predictive errors, with negligible differences in NRMSE. 
Figure \ref{fig:time_approx_error_unormalized_Vicsek}(c) shows a comparison between the log-likelihood values computed via the direct computation and the IKF-CG algorithm, where the latter employs the log-determinant approximation detailed in the Supplementary Material. 
Both methods produce similar log-likelihood values across different sample sizes.

\begin{figure}[tp] 
\centering 
\includegraphics[width=15cm]{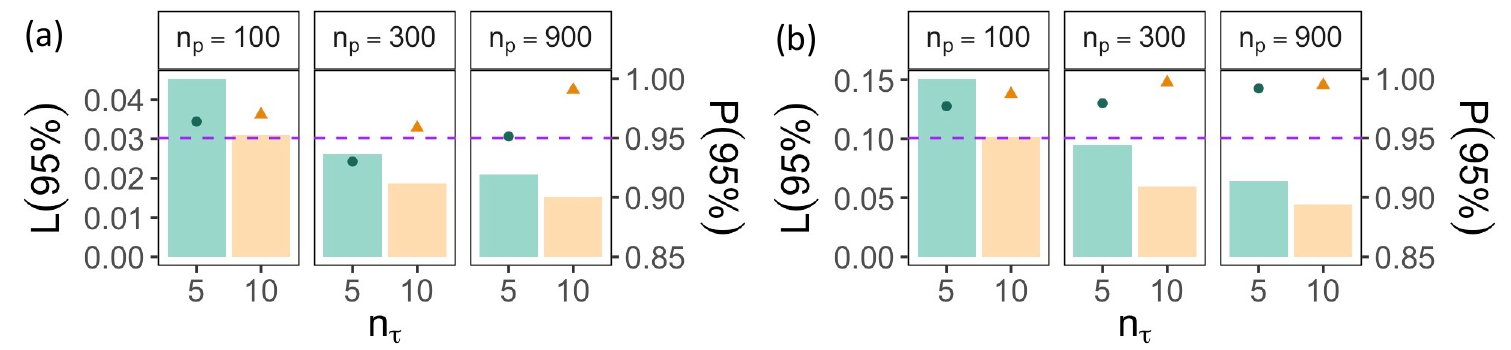}
\caption{
Uncertainty assessment of predicted interaction function in the unnormalized Vicsek model with $\sigma_0^2 = 0.1^2$ using (a) Mat{\'e}rn covariance function in (\ref{equ:matern_5_2}) and (b) exponential covariance function. Bars represent average lengths of 95\% posterior credible intervals (\ref{equ:length}). Dots indicate the proportions covered in the 95\% intervals (\ref{equ:coverage}), with the optimal coverage (0.95) shown as the purple dashed lines. 
}
\label{fig:unnormalized_Vicsek_interval_0.1}
\end{figure}

\begin{figure}[tp] 
\centering
    \includegraphics[width=16cm]{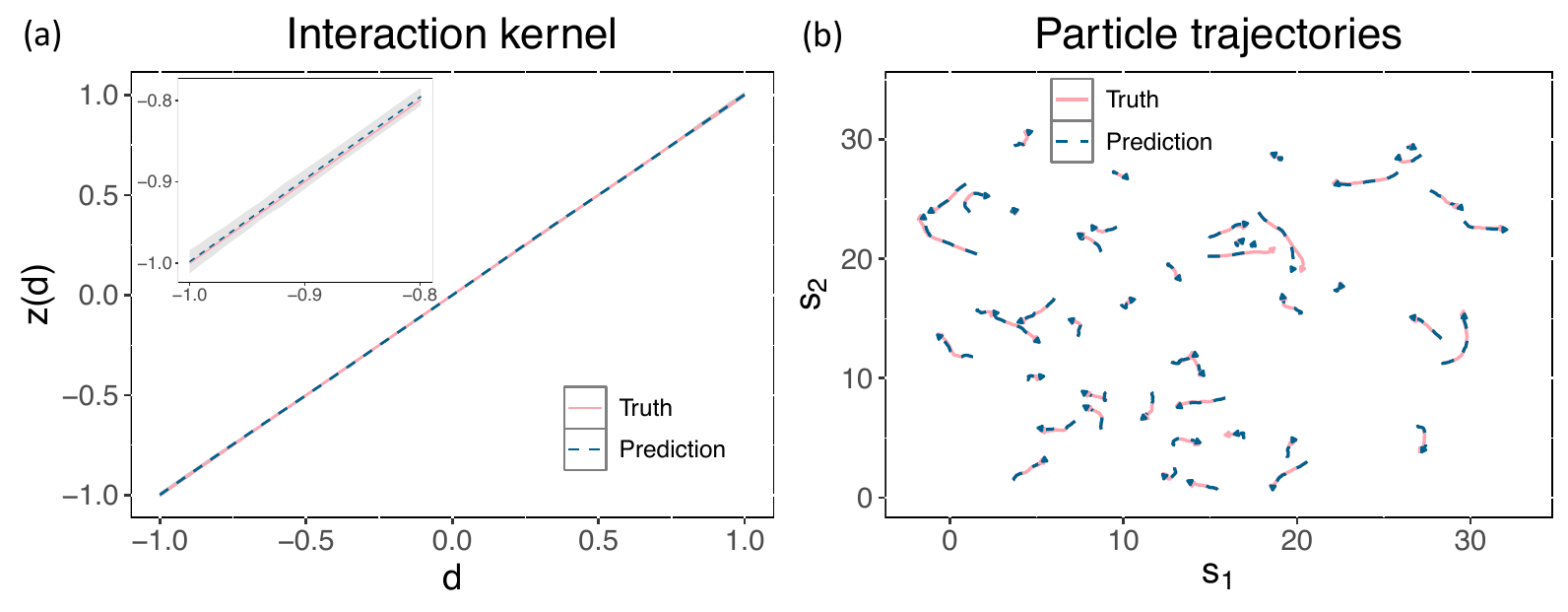}
    \caption{{(a) Predicted (blue dashed lines) and true (pink solid lines) interaction functions with the $95\%$ posterior credible interval (grey shaded area) for the unnormalized Vicsek model with $n_{\tau}=900$, $T=10$, and $\sigma_0^2=0.1^2$, using Mat{\'e}rn covariance function with $\nu=2.5$. (b) Trajectories of $45$ randomly sampled particles over $50$ time frames using estimated (blue dashed lines) and true (pink solid lines) interaction functions with identical initial positions and noise samples.}}
\label{fig:kernel_trajectory}
\end{figure}

Next, we evaluate the performance of the IKF-CG algorithm across 12 scenarios with varying particle numbers ($n_{p}=100, 300, 900$), time frames ($n_{\tau}=5, 10$), and noise variances ($\sigma^2_0=0.1^2, 0.2^2$). The predictive performance is evaluated using $n^*=200$ test velocity inputs evenly spaced across a representative input domain of the interaction function in the data, $[-1,1]$, averaged over $E=20$ experiments. 

Panels (a) and (b) of Figs. \ref{fig:unnormalized_Vicsek_nrmse_0.1}-\ref{fig:unnormalized_Vicsek_interval_0.1} show the predictive performance of particle interactions using the Mat{\'e}rn covariance with roughness parameter $\nu=2.5$ and the exponential covariance (Mat{\'e}rn covariance with $\nu=0.5$), respectively, for noise variance $\sigma_0^2=0.1^2$. Results for $\sigma_0^2=0.2^2$ are provided in the Supplementary Material. Across all scenarios, NRMSEs remain low, with improved accuracy for larger datasets and smaller noise variances. 
The decrease in the average length of the $95\%$ posterior credible interval with increasing sample size, along with the relatively small interval span compared to the output range, indicates improved prediction confidence with more observations. Moreover, the coverage proportion for the $95\%$ posterior credible interval is close to the nominal $95\%$ level in all cases, validating the reliability of the uncertainty quantification. 
By comparing panels (a) and (b) in Figs. \ref{fig:unnormalized_Vicsek_nrmse_0.1}-\ref{fig:unnormalized_Vicsek_interval_0.1}, we find that the 
Mat{\'e}rn covariance in  (\ref{equ:matern_5_2}) 
yields lower NRMSE and narrower $95\%$ posterior credible intervals than the exponential kernel. This improvement is due to the smoother latent process induced by (\ref{equ:matern_5_2}), which is twice mean-squared differentiable, while the process with an exponential kernel is not differentiable.

Figure \ref{fig:kernel_trajectory}(a) shows close agreement between predicted and true interaction functions. 
The $95\%$ posterior credible interval is narrow  
yet covers approximately 95\% of the test samples of the interaction function. 
Figure \ref{fig:kernel_trajectory}(b) compares particle trajectories over $50$ time steps by the predicted mean of one-step-ahead predictions and the true interaction function, both having the same noise vectors. The trajectories are visually similar. 
}
	
\subsection{A modified Vicsek model with multiple interactions}
\label{sec:modify_Vicsek}
Various extensions of the Vicsek model have been studied  
to capture more complex particle dynamics \cite{chate2008modeling,ginelli2010large}. For illustration, we consider a modified Vicsek model with two interaction functions. The 2-dimensional velocity $\mathbold v_{i'}(\tau) = (v_{i',1}(\tau), v_{i',2}(\tau))^T$, corresponding to the output in (\ref{equ:dynamic_factors}), is updated according to: 
\begin{align}
      \mathbold  v_{i'}(\tau)=\frac{\sum_{k \in ne_{i'}(\tau-1)}  \mathbold  v_{k}(\tau-1) }{p_{i'}(\tau-1)}+\frac{\sum_{k\in ne^{'}_{i'}(\tau-1)    }  f(d_{i',k}(\tau-1))   \mathbold  e_{i',k}(\tau-1)  }{p'_{i'}(\tau-1)} +  \bmbold  \epsilon_{i'}(\tau),  
    \label{equ:modified_vicsek}
\end{align}
where $\bmbold  \epsilon_{i'}(\tau)=(\epsilon_{i',1}(\tau),\epsilon_{i',2}(\tau))^T$ is a Gaussian noise vector with variance $\sigma_0^2$ and $D_y=2$. 
The first term in (\ref{equ:modified_vicsek}) models velocity alignment with neighboring particles, and the second term introduces a distance-dependent interaction function $f(\cdot)$. The definitions of neighbor sets, 2-dimensional vector $\mathbold e_{i',k}(\tau-1)$, and 
the second interaction function $f(\cdot)$ are provided in Section S8.2 of the Supplementary Material.

We simulate 12 scenarios, each replicated $E=20$ times, using the same number of particles, time frame, and noise level as in the unnormalized Vicsek model in Section \ref{sec:Vicsek}. 
The predictive performance of the IKF-CG algorithm is evaluated using $n^* = 200$ test inputs evenly spaced across the input 
domain of each interaction function. We focus on the model with the Mat{\'e}rn kernel in (\ref{equ:matern_5_2}), with the results for the exponential kernel reported in the Supplementary Material. 
Consistent with the results of the unnormalized Vicsek model, we find that models with the Mat{\'e}rn kernel in (\ref{equ:matern_5_2}) are more accurate due to 
the smooth prior imposed on the interaction function. 

\begin{table}[t]
\centering
\begin{tabular}{cccccccc}
&  & \multicolumn{3}{c}{First Interaction}  & \multicolumn{3}{c}{Second Interaction}\\ 
& & \multicolumn{1}{c}{NRMSE} & \multicolumn{1}{c}{$L_1(95\%)$} & \multicolumn{1}{c}{$P_1(95\%)$} & \multicolumn{1}{c}{NRMSE} & \multicolumn{1}{c}{$L_2(95\%)$} & $ P_2(95\%)$ \\ 
\multicolumn{1}{c}{$n_{p}=100$}  &$n_{\tau}=5$  &\multicolumn{1}{c}{$6.8\times 10^{-3}$}  &\multicolumn{1}{c}{$0.098$}  &\multicolumn{1}{c}{$92\%$}  &\multicolumn{1}{c}{$8.7\times 10^{-2}$}  &\multicolumn{1}{c}{$0.29$}  &$94\%$ \\
\multicolumn{1}{c}{}           &$n_{\tau}=10$ &\multicolumn{1}{c}{$6.0\times 10^{-3}$}  &\multicolumn{1}{c}{$0.097$}  &\multicolumn{1}{c}{$95\%$}  &\multicolumn{1}{c}{$3.8\times 10^{-2}$}  &\multicolumn{1}{c}{$0.18$}  &$96\%$ \\
\multicolumn{1}{c}{$n_{p}=300$} & $n_{\tau}=5$  & \multicolumn{1}{c}{$1.2\times 10^{-2}$}      & \multicolumn{1}{c}{$0.23$}   &  \multicolumn{1}{c}{$87\%$} & \multicolumn{1}{c}{$3.2\times 10^{-2}$}      & \multicolumn{1}{c}{$0.21$}       & $96\%$ \\
\multicolumn{1}{c}{}      & $n_{\tau}=10$ & \multicolumn{1}{c}{$4.2\times 10^{-3}$}      & \multicolumn{1}{c}{$0.10$}       & \multicolumn{1}{c}{$98\%$}  & \multicolumn{1}{c}{$2.0\times 10^{-2}$}      & \multicolumn{1}{c}{$0.13$}       &  $97\%$ \\
\multicolumn{1}{c}{$n_{p}=900$} & $n_{\tau}=5$  & \multicolumn{1}{c}{$5.6\times 10^{-3}$}      & \multicolumn{1}{c}{$0.10$}       & \multicolumn{1}{c}{$96\%$}  & \multicolumn{1}{c}{$1.4\times 10^{-2}$}      & \multicolumn{1}{c}{$0.12$}       & $98\%$  \\
\multicolumn{1}{c}{}      & $n_{\tau}=10$ & \multicolumn{1}{c}{$4.4\times 10^{-3}$}      & \multicolumn{1}{c}{$0.12$}       &  \multicolumn{1}{c}{$97\%$}  & \multicolumn{1}{c}{$1.4\times 10^{-2}$}      & \multicolumn{1}{c}{$0.10$}       &   $96\%$  \\ 
\end{tabular}
\caption{The predictive accuracy and uncertainty assessment by (\ref{equ:nrmse})-(\ref{equ:coverage}) for the modified Vicsek model with $\sigma_0^2 = 0.1^2$ using Mat{\'e}rn covariance function and roughness parameter $\nu_j = 2.5$.}
\label{tab:vicsek_modified_matern_5_2}
\end{table}

Table \ref{tab:vicsek_modified_matern_5_2} summarizes the predictive performance for $\sigma_0^2=0.1^2$; results for $\sigma_0^2=0.2^2$ are reported in the Supplementary Material. 
While both interaction functions exhibit relatively low NRMSEs, 
the second interaction function has a higher NRMSE than the first, 
because the repulsion at short distances creates fewer training samples with close-proximity neighbors. This discrepancy can be mitigated by increasing the number of observations. 
The average length of the $95\%$  posterior credible interval for the second interaction decreases as the sample size increases, and the coverage proportion remains close to the nominal 95\% level across all scenarios.

\begin{figure}[t!] 
\centering
    \includegraphics[width=16cm]{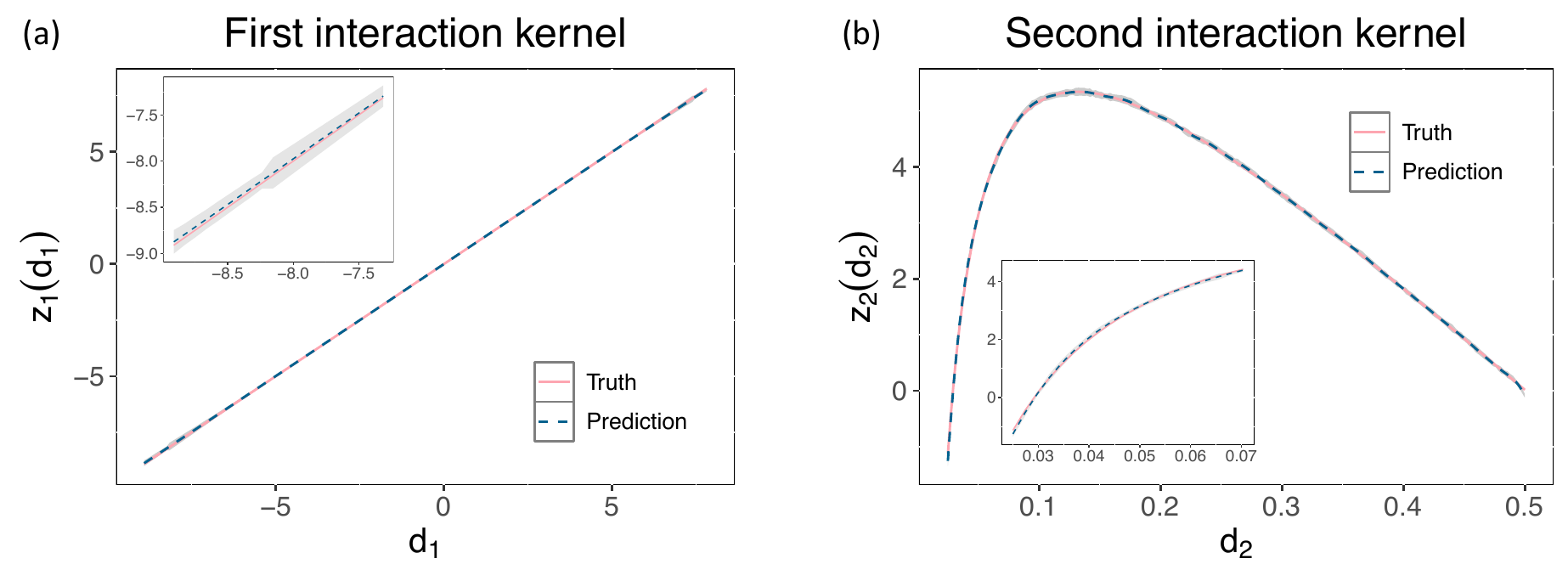}
   \caption{Predictions (blue dashed curves), truth (pink solid curves), and the $95\%$ posterior credible interval (shaded regions) of particle interaction functions when $n_{p}=900$, $n_{\tau}=10$, and $\sigma_0^2=0.1^2$. 
  }
\label{fig:Vicsek_variation_kernels}
\end{figure}

In Fig. \ref{fig:Vicsek_variation_kernels}, we show the predictions of the interaction functions, where the shaded regions represent the 95\% posterior credible intervals. The input velocities are larger than the unnormalized Vicsek model due to the inclusion of the second interaction function, and thus the test input domain in  Fig. \ref{fig:Vicsek_variation_kernels}(a) is wider than that in Fig. \ref{fig:kernel_trajectory}(a).  The predictions closely match the true values, and the credible intervals, while narrow relative to the output range and nearly invisible in the plots, mostly cover the truth. These results suggest high confidence in the predictions.

\subsection{Estimating cell-cell interactions on  anisotropic substrates}
\label{sec:cell} 
We analyze video microscopy data from \cite{luo2023molecular}, which tracks the trajectories of over 2,000 human dermal fibroblasts moving on a nematically aligned, liquid-crystalline substrate. 
This experimental setup encourages cellular alignment along a horizontal axis, with alignment order increasing over time, though the underlying mechanism remains largely unknown. 
Cells were imaged every 20 minutes over a 36-hour period, 
during which the cell count grew from 2,615 to 2,953 due to proliferation. 
Our objective is to estimate the latent interaction function between cells. Given the vast number of velocity observations ($\tilde N\approx 300,000$), direct formation and inversion of the covariance matrix is impractical.

We apply the IKF-CG algorithm to estimate the latent interaction function. Because of the anisotropic substrate, cellular velocities differ between horizontal and vertical directions, so we independently model the velocity of the $i'$th cell in each direction $l$ as
\begin{equation}
v_{i',l}(\tau)=\frac{1}{p_{ne_{i',l}(\tau-1)}}\sum_{k \in ne_{i',l}(\tau-1)} z_l(d_{k,l})+\epsilon_{i',l}(\tau), \quad i'=1,\dots,n_{p}(\tau)
\label{equ:v_i_prime_l}
\end{equation}
where $n_{p}(\tau)$ is the cell count at time $\tau$, $l=1,2$ correspond to horizontal and vertical directions, respectively, and $\epsilon_{i',l}(\tau)\sim \mathcal N(0,\sigma^2_{0,l}(\tau))$ denotes the noise. Inspired by the modified Vicsek model, we set $d_{k,l}=v_{k,l}(\tau-1)$. To account for velocity decay caused by increasing cell confluence, we model the noise variances as $\sigma^2_{0,l}(\tau)=\omega_l\sigma^2_{v,l}(\tau-1)$, where $\sigma^2_{v,l}(\tau-1)$ is the sample velocity variance at time $(\tau-1)$,  and $\omega_l$ is a parameter estimated from data for $l=1,2$. 
The neighboring set in (\ref{equ:v_i_prime_l}), formed as $ne_{i',l}(\tau-1)=\{k: ||\mathbold s_{i'}(\tau-1)-  \mathbold s_{k}(\tau-1)||<r_l \mbox{ and } \mathbold v_{i'}(\tau-1)^T \mathbold v_{k}(\tau-1)>0\}$, where $r_l$ denotes the interaction radius of direction $l$ for $l=1,2$,  excludes particles moving in opposite directions, reflecting the observed gliding and intercalating behavior of cells \cite{walck2014cell}.

\begin{figure}[t] 
\centering
\includegraphics[width=15cm]{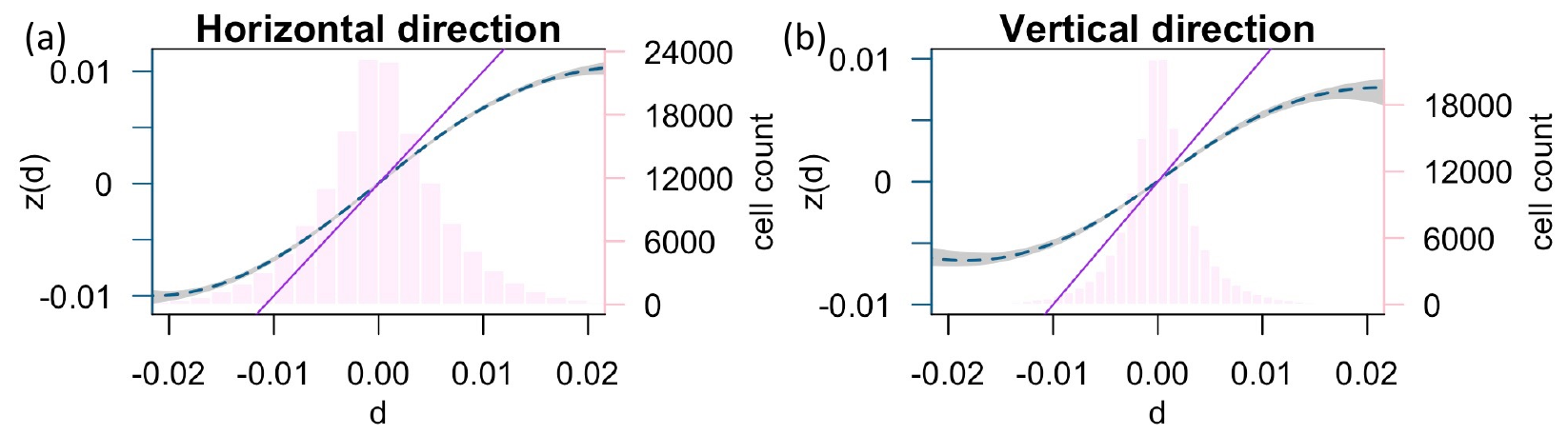}
\caption{The blue dashed curves show the predicted interaction function for the horizontal and vertical directions. The grey shaded area is the 95\% posterior credible interval. The purple line of slope 1 is the prediction using the unnormalized Vicsek model. The light pink histogram shows the velocity distribution of all training samples. 
}
\label{fig:boostrap_interval_cell}
\end{figure}

We use observations from the first half of the time frames as the training set and the latter half as the testing set. The predicted interaction functions in Fig. \ref{fig:boostrap_interval_cell} show diminishing effects in both directions, likely due to cell-substrate interactions such as friction. 
The estimated uncertainty (obtained via residual bootstrap) 
increases when the absolute velocity of neighboring cells in the previous time frame is large, which is attributed to fewer observations at the boundaries of the input domain. Furthermore, the interaction in the vertical direction is weaker than in the horizontal direction, due to the confinement from the anisotropic substrate in the vertical direction.

{Our nonparametric estimation of the interaction functions is compared with two models for one-step-ahead forecasts of directional velocities: the unnormalized Vicsek model introduced in Section \ref{sec:Vicsek}, which uses a shared interaction radius for both velocity components, 
and the anisotropic Vicsek model, which uses distinct radii in each direction.  
As shown in Table \ref{tab:pred_cell}, both Vicsek models have similar RMSE as the difference of interaction radii does not have a large impact for estimating interaction functions.  Our model outperforms both Vicsek models in the RMSE for one-step-ahead forecasts of directional velocities, despite the large inherent stochasticity in cellular motion. 
Moreover, our model has notably shorter average interval lengths that cover approximately $95\%$ of the held-out observations. These findings underscore the importance of the IFK-CG algorithm in enabling the use of large datasets to overcome high stochasticity and capture the underlying dynamics of cell alignment processes.}

\begin{table}[t]
\centering

\begin{tabular}{ccccccc}
\multicolumn{1}{c}{} & \multicolumn{3}{c}{Horizontal direction}    & \multicolumn{3}{c}{Vertical direction}   \\
\multicolumn{1}{c}{} & \multicolumn{1}{c}{RMSE} & \multicolumn{1}{c}{L(95\%)} & \multicolumn{1}{c}{P(95\%)} & \multicolumn{1}{c}{RMSE} & \multicolumn{1}{c}{L(95\%)} & \multicolumn{1}{c}{P(95\%)} \\
Unnormalized Vicsek             & $3.9\times 10^{-3}$          & 0.029                    & 99\%                      & $2.6\times 10^{-3}$          & 0.029                  & 99\%    \\
Anisotropic Vicsek              & $3.9\times 10^{-3}$          & 0.034                    & 99\%                      & $2.6\times 10^{-3}$          & 0.022                    & 99\%    \\
Nonparametric estimation            & $3.5\times 10^{-3}$            & 0.015                    & 95\%                      & $2.3\times 10^{-3}$            & 0.0092                    & 95\%                      \\
\end{tabular}
\caption{One-step-ahead prediction performance on the testing dataset. Here  $\mbox{RMSE}=\{\sum_{\tau=1}^{n_{\tau}^*}\sum_{i'=1}^{n_{p}(\tau)}(\hat v_{i',l}(\tau)-v_{i',l}(\tau))^2/\sum_{\tau=1}^{n_{\tau}^*}n_{p}(\tau)\}^{1/2}$, with $n_{\tau}^*$ denoting the number of testing time frames. L(95\%) and P(95\%) are computed similarly to those in (\ref{equ:length}) - 
(\ref{equ:coverage}), but with the predictive interval of $z$ replaced by that of $y$, where $y$ represents the velocity in each direction.}
\label{tab:pred_cell}
\end{table}

\section{Concluding remarks}
\label{sec:conclusion}
Several future directions are worth exploring. First,  computing large matrix-vector products is ubiquitous, and the approach can be extended to different models, including some large neural network models. 
Second, it is an open topic to scalably compute the logarithm of the determinant of the covariance in (\ref{equ:Sigma_y}) without approximation.  
Third, the new approach can be extended to estimate latent functions when forward equations are unavailable or computationally intensive in nonlinear or non-Gaussian dynamical systems, where the ensemble Kalman filter \cite{evensen2009data,stroud2010ensemble} and particle filter \cite{kitagawa1996monte} are commonly used. 
Fourth, our  approach can be extended for estimating and predicting 
multivariate time series \cite{lam2011estimation} and generalized factor processes for categorical observations  \cite{durante2014nonparametric}. {Finally, we aim to generalize the fast approach to scenarios with input dimensions larger than one.}  

\section*{Acknowledgement}
Xinyi Fang acknowledges partial support from the BioPACIFIC Materials Innovation Platform of the National Science Foundation under Award No. DMR-1933487. Mengyang Gu acknowledges partial support from the National Science Foundation under Award No. OAC-2411043 and the University of California  Multicampus Research Programs and Initiatives under Award No. M23PL599. We thank the editor, associate editor, and three anonymous referees for their comments that substantially improved this article. 
 
\section*{Supplementary Material}
The Supplementary Material contains (i) proofs of all lemmas; (ii) a summary of the connection between Gaussian processes with Mat{\'e}rn covariance (roughness parameters 0.5 and 2.5) and DLMs, with closed-form expressions of $  \mathbold  F_{t,j}$,  $  \mathbold  G_{t,j}$ and $  \mathbold  W_{t,j}$; (iii) the conjugate gradient algorithm; (iv) procedures for the scalable computation of the predictive distribution in particle dynamics; (v) an application of the IKF-CG algorithm for predicting missing values in lattice data; (vi)  parameter estimation methods; (vii)  computational complexity analysis; (viii) additional simulation results for estimating particle interaction; and (ix) numerical results for predicting missing values in lattice data.

\appendix

\section*{Appendix A. Kalman Filter}
\label{sec:Appendix_KF}

\newtheorem{apxlemma}{Lemma}[section]
\renewcommand{\theapxlemma}{A\arabic{apxlemma}}

\renewcommand{\theequation}{A\arabic{equation}}
\setcounter{equation}{0}

The Kalman filter \cite{kalman1960new} for dynamic linear models in (\ref{equ:DLM_y})-(\ref{equ:DLM_theta}) is summarized in  Lemma \ref{lemma:KF}. 
\begin{apxlemma} 
Let $  \bmbold  \theta_{t-1}\mid {  \mathbold  y}_{1:t-1} \sim  \mathcal{MN}(   \mathbold  m_{t-1},   \mathbold  C_{t-1})$. Recursively for $t=2,\dots,{N}$ the following statements hold.
\begin{itemize}
\item[(i)] The one-step-ahead predictive distribution of $  \bmbold  \theta_t$ given  ${  \mathbold  y}_{1:t-1}$ is 
 \begin{equation}
   \bmbold  \theta_t \mid {  \mathbold  y}_{1:t-1} \sim  \mathcal{MN}(  \mathbold  b_t,   \mathbold  B_t),  
 \label{equ:KF1}
 \end{equation}
with $  \mathbold  b_t=   \mathbold  G_t   \mathbold  m_{t-1} $ and $  \mathbold  B_t=   \mathbold  G_t   \mathbold  C_{t-1}   \mathbold  G^T_t+  \mathbold  W_t$.  
\item[(ii)] The one-step-ahead predictive distribution of $y_t$ given ${  \mathbold  y}_{1:t-1}$ is  
 \begin{equation}
y_t\mid {  \mathbold  y}_{1:t-1} \sim  \mathcal{N}(f_t, Q_t), 
 \label{equ:KF2}
\end{equation}
with   $f_t=   \mathbold  F_t   \mathbold  b_{t}, $ and $Q_t=   \mathbold  F_t   \mathbold  B_t   \mathbold  F^T_t+  V_t$. 
\item[(iii)] The filtering distribution of $  \bmbold  \theta_t$ given ${  \mathbold  y}_{1:t}$ follows  
 \begin{equation}
  \bmbold  \theta_t\mid {  \mathbold  y}_{1:t} \sim  \mathcal{MN}(  \mathbold  m_t,   \mathbold  C_t),  
 \label{equ:KF3}
\end{equation}
with  $  \mathbold  m_t=   \mathbold  b_t +   \mathbold  B_t    \mathbold  F^T_t  Q^{-1}_t  (y_t -f_t)$ and  $  \mathbold  C_t=   \mathbold  B_t -    \mathbold  B_t   \mathbold  F^T_t   Q^{-1}_t   \mathbold  F_t   \mathbold  B_t$. 
\end{itemize}
\label{lemma:KF}
\end{apxlemma}

 \begin{apxlemma}
Define $  \mathbold  {\tilde y}= (\tilde y_{1},\dots, \tilde y_{N})^T=  \mathbold  {L}^{-1}   \mathbold  {y} $, where $  \mathbold  {L}$ is the  factor in Cholesky decomposition of a positive definite covariance  ${  \bmbold  \Sigma}$, with ${  \bmbold  \Sigma}= \mbox{cov}[  \mathbold  {y}
]$ and $  \mathbold  {y} = (y_{1},\dots, y_{N})^T$ defined as in (\ref{equ:DLM_y}). We have  
  \begin{equation}
   \tilde y_{t}=\frac{y_t -f_t}{ {Q}^{1/2}_t},
   \label{equ:L_inv_y} 
   \end{equation}
  where $f_t$ and $Q_t$ are defined in (\ref{equ:KF2}). Furthermore, the $t$th diagonal term of $  \mathbold  {L}$ is $    L_{t,t}={Q}^{1/2}_t$. 
  \label{lemma:L_inv_z}
 \end{apxlemma}

\bibliographystyle{plain}
\bibliography{References_chronical_2022}

@article{hestenes1952methods,
  title={Methods of conjugate gradients for solving linear systems},
  author={Hestenes, Magnus R and Stiefel, Eduard},
  journal={Journal of research of the National Bureau of Standards},
  volume={49},
  number={6},
  pages={409},
  year={1952}
}

@article{whittle1954stationary,
  title={On stationary processes in the plane},
  author={Whittle, Peter},
  journal={Biometrika},
  pages={434--449},
  year={1954},
  publisher={JSTOR}
}

@article{kalman1960new,
  title={A new approach to linear filtering and prediction problems},
  author={Kalman, Rudolph Emil},
  journal={Journal of basic Engineering},
  volume={82},
  number={1},
  pages={35--45},
  year={1960},
  publisher={American Society of Mechanical Engineers}
}

@article{vecchia1988estimation,
  title={Estimation and model identification for continuous spatial processes},
  author={Vecchia, Aldo V},
  journal={Journal of the Royal Statistical Society: Series B (Methodological)},
  volume={50},
  number={2},
  pages={297--312},
  year={1988},
  publisher={Wiley Online Library}
}

@article{handcock1993bayesian,
  title={A {B}ayesian analysis of kriging},
  author={Handcock, Mark S and Stein, Michael L},
  journal={Technometrics},
  volume={35},
  number={4},
  pages={403--410},
  year={1993},
  publisher={Taylor \& Francis Group}
}

@article{hastie1993varying,
  title={Varying-coefficient models},
  author={Hastie, Trevor and Tibshirani, Robert},
  journal={Journal of the Royal Statistical Society: Series B (Methodological)},
  volume={55},
  number={4},
  pages={757--779},
  year={1993},
  publisher={Wiley Online Library}
}

@article{vicsek1995novel,
  title={Novel type of phase transition in a system of self-driven particles},
  author={Vicsek, Tam{\'a}s and Czir{\'o}k, Andr{\'a}s and Ben-Jacob, Eshel and Cohen, Inon and Shochet, Ofer},
  journal={Physical review letters},
  volume={75},
  number={6},
  pages={1226},
  year={1995},
  publisher={APS}
}

@article{kitagawa1996monte,
  title={Monte {C}arlo filter and smoother for non-{G}aussian nonlinear state space models},
  author={Kitagawa, Genshiro},
  journal={Journal of computational and graphical statistics},
  volume={5},
  number={1},
  pages={1--25},
  year={1996},
  publisher={Taylor \& Francis}
}

@BOOK{West1997,
  title = {Bayesian Forecasting \& Dynamic Models},
  publisher = {Springer Verlag},
  year = {1997},
  author = {M. West and P. J. Harrison},
  edition = {2nd},
  key = {WestHarrison.YellowBook.1997},
  url = {http://www.stat.duke.edu/~mw/book.html}
}

@book{rapaport2004art,
  title={The art of molecular dynamics simulation},
  author={Rapaport, Dennis C},
  year={2004},
  publisher={Cambridge university press}
}

@article{snelson2006sparse,
  title={Sparse {G}aussian processes using pseudo-inputs},
  author={Snelson, Edward and Ghahramani, Zoubin},
  journal={Advances in neural information processing systems},
  volume={18},
  pages={1257},
  year={2006},
  publisher={Citeseer}
}

@article{furrer2006covariance,
  title={Covariance tapering for interpolation of large spatial datasets},
  author={Furrer, Reinhard and Genton, Marc G and Nychka, Douglas},
  journal={Journal of Computational and Graphical Statistics},
  volume={15},
  number={3},
  pages={502--523},
  year={2006},
  publisher={Taylor \& Francis}
}

@article{chate2008modeling,
  title={Modeling collective motion: variations on the {V}icsek model},
  author={Chat{\'e}, Hugues and Ginelli, Francesco and Gr{\'e}goire, Guillaume and Peruani, Fernando and Raynaud, Franck},
  journal={The European Physical Journal B},
  volume={64},
  number={3},
  pages={451--456},
  year={2008},
  publisher={Springer}
}

@article{cressie2008fixed,
  title={Fixed rank kriging for very large spatial data sets},
  author={Cressie, Noel and Johannesson, Gardar},
  journal={Journal of the Royal Statistical Society: Series B (Statistical Methodology)},
  volume={70},
  number={1},
  pages={209--226},
  year={2008},
  publisher={Wiley Online Library}
}

@book{evensen2009data,
  title={Data assimilation: the ensemble Kalman filter},
  author={Evensen, Geir},
  year={2009},
  publisher={Springer Science \& Business Media}
}

@incollection{petris2009dynamic,
  title={Dynamic linear models},
  author={Petris, Giovanni and Petrone, Sonia and Campagnoli, Patrizia},
  booktitle={Dynamic linear models with R},
  year={2009},
  publisher={Springer}
}

@article{katz2011inferring,
  title={Inferring the structure and dynamics of interactions in schooling fish},
  author={Katz, Yael and Tunstr{\o}m, Kolbj{\o}rn and Ioannou, Christos C and Huepe, Cristi{\'a}n and Couzin, Iain D},
  journal={Proceedings of the National Academy of Sciences},
  volume={108},
  number={46},
  pages={18720--18725},
  year={2011},
  publisher={National Acad Sciences}
}

@book{prado2010time,
  title={Time series: modeling, computation, and inference},
  author={Prado, Raquel and West, Mike},
  year={2010},
  publisher={Chapman and Hall/CRC}
}

@article{ginelli2010large,
  title={Large-scale collective properties of self-propelled rods},
  author={Ginelli, Francesco and Peruani, Fernando and B{\"a}r, Markus and Chat{\'e}, Hugues},
  journal={Physical review letters},
  volume={104},
  number={18},
  pages={184502},
  year={2010},
  publisher={APS}
}

@article{lindgren2011explicit,
  title={An explicit link between {G}aussian fields and {G}aussian {M}arkov random fields: the stochastic partial differential equation approach},
  author={Lindgren, Finn and Rue, H{\aa}vard and Lindstr{\"o}m, Johan},
  journal={Journal of the Royal Statistical Society: Series B (Statistical Methodology)},
  volume={73},
  number={4},
  pages={423--498},
  year={2011},
  publisher={Wiley Online Library}
}

@inproceedings{hartikainen2010kalman,
  title={Kalman filtering and smoothing solutions to temporal {G}aussian process regression models},
  author={Hartikainen, Jouni and Sarkka, Simo},
  booktitle={Machine Learning for Signal Processing (MLSP), 2010 IEEE International Workshop on},
  pages={379--384},
  year={2010},
  organization={IEEE}
}

@article{stroud2010ensemble,
  title={An ensemble {K}alman filter and smoother for satellite data assimilation},
  author={Stroud, Jonathan R and Stein, Michael L and Lesht, Barry M and Schwab, David J and Beletsky, Dmitry},
  journal={Journal of the american statistical association},
  volume={105},
  number={491},
  pages={978--990},
  year={2010},
  publisher={Taylor \& Francis}
}

@article{lam2011estimation,
  title={Estimation of latent factors for high-dimensional time series},
  author={Lam, Clifford and Yao, Qiwei and Bathia, Neil},
  journal={Biometrika},
  volume={98},
  number={4},
  pages={901--918},
  year={2011},
  publisher={Oxford University Press}
}

@book{durbin2012time,
  title={Time series analysis by state space methods},
  author={Durbin, James and Koopman, Siem Jan},
  volume={38},
  year={2012},
  publisher={OUP Oxford}
}

@book{stein2012interpolation,
  title={Interpolation of spatial data: some theory for kriging},
  author={Stein, Michael L},
  year={2012},
  publisher={Springer Science \& Business Media}
}

@article{stein2013stochastic,
  title={Stochastic approximation of score functions for {G}aussian processes},
  author={Stein, Michael L and Chen, Jie and Anitescu, Mihai},
  journal={The Annals of Applied Statistics},
  volume={7},
  number={2},
  pages={1162--1191},
  year={2013},
  publisher={Institute of Mathematical Statistics}
}

@article{durante2014nonparametric,
  title={Nonparametric {B}ayes dynamic modelling of relational data},
  author={Durante, Daniele and Dunson, David B},
  journal={Biometrika},
  volume={101},
  number={4},
  pages={883--898},
  year={2014},
  publisher={Oxford University Press}
}

@article{walck2014cell,
  title={Cell intercalation from top to bottom},
  author={Walck-Shannon, Elise and Hardin, Jeff},
  journal={{Nature Reviews Molecular Cell Biology}},
  volume={15},
  number={1},
  pages={34--48},
  year={2014},
  publisher={Nature Publishing Group UK London}
}

@article{gramacy2015local,
  title={Local {G}aussian process approximation for large computer experiments},
  author={Gramacy, Robert B and Apley, Daniel W},
  journal={Journal of Computational and Graphical Statistics},
  volume={24},
  number={2},
  pages={561--578},
  year={2015},
  publisher={Taylor \& Francis}
}

@article{datta2016hierarchical,
  title={Hierarchical nearest-neighbor {G}aussian process models for large geostatistical datasets},
  author={Datta, Abhirup and Banerjee, Sudipto and Finley, Andrew O and Gelfand, Alan E},
  journal={Journal of the American Statistical Association},
  volume={111},
  number={514},
  pages={800--812},
  year={2016},
  publisher={Taylor \& Francis}
}

@article{saibaba2017randomized,
  title={Randomized matrix-free trace and log-determinant estimators},
  author={Saibaba, Arvind K and Alexanderian, Alen and Ipsen, Ilse CF},
  journal={Numerische Mathematik},
  volume={137},
  number={2},
  pages={353--395},
  year={2017},
  publisher={Springer}
}

@article{Gu2018robustness,
  title={Robust {G}aussian stochastic process emulation},
  author={Gu, Mengyang and Wang, Xiaojing and Berger, James O},
    journal={Annals of Statistics},
      volume={46},
  number={6A},
  pages={3038--3066},
    year    = {2018},

}

@article{lu2019nonparametric,
  title={Nonparametric inference of interaction laws in systems of agents from trajectory data},
  author={Lu, Fei and Zhong, Ming and Tang, Sui and Maggioni, Mauro},
  journal={Proceedings of the National Academy of Sciences},
  volume={116},
  number={29},
  pages={14424--14433},
  year={2019},
  publisher={National Acad Sciences}
}

@Manual{gu2025fastgasppackage,
    title = {FastGaSP: Fast and Exact Computation of {G}aussian Stochastic Process},
    author = {Gu, Mengyang and  Fang, Xinyi and Lin, Yizi},
    year = {2025},
    note = {R package version 0.6.1},
    url = {https://CRAN.R-project.org/package=FastGaSP},
  }

@article{katzfuss2022scaled,
  title={Scaled {V}ecchia approximation for fast computer-model emulation},
  author={Katzfuss, Matthias and Guinness, Joseph and Lawrence, Earl},
  journal={SIAM/ASA Journal on Uncertainty Quantification},
  volume={10},
  number={2},
  pages={537--554},
  year={2022},
  publisher={SIAM}
}

@article{katzfuss2021general,
  title={A general framework for {V}ecchia approximations of {G}aussian processes},
  author={Katzfuss, Matthias and Guinness, Joseph},
  journal={Statistical Science},
  volume={36},
  number={1},
  pages={124--141},
  year={2021},
  publisher={Institute of Mathematical Statistics}
}

@article{finley2022spnngp,
 title={{spNNGP R Package for Nearest Neighbor Gaussian Process Models}},
 volume={103},
 url={https://www.jstatsoft.org/index.php/jss/article/view/v103i05},
 doi={10.18637/jss.v103.i05},
 number={5},
 journal={Journal of Statistical Software},
 author={Finley, Andrew O. and Datta, Abhirup and Banerjee, Sudipto},
 year={2022},
 pages={1–40}
}

@article{luo2023molecular,
  title={Molecular-scale substrate anisotropy, crowding and division drive collective behaviours in cell monolayers},
  author={Luo, Yimin and Gu, Mengyang and Park, Minwook and Fang, Xinyi and Kwon, Younghoon and Urue{\~n}a, Juan Manuel and Read de Alaniz, Javier and Helgeson, Matthew E and Marchetti, Cristina M and Valentine, Megan T},
  journal={Journal of the Royal Society Interface},
  volume={20},
  number={204},
  pages={20230160},
  year={2023},
  publisher={The Royal Society}
}

@article{majumder2022kryging,
  title={Kryging: geostatistical analysis of large-scale datasets using {K}rylov subspace methods},
  author={Majumder, Suman and Guan, Yawen and Reich, Brian J and Saibaba, Arvind K},
  journal={Statistics and Computing},
  volume={32},
  number={5},
  pages={74},
  year={2022},
  publisher={Springer}
}

\end{document}